# Expansion of landing areas on the Venus surface using resonant orbits in the Venera-D project

Natan Eismont[a], Vladislav Zubko[a,c], Andrey Belyaev[a,c], Konstantin Fedyaev[a,d], Lyudmila Zasova[a], Dmitry Gorinov[a], Alexander Simonov[b], Ravil Nazirov[a]

[a]*Department of Space Dynamics and Mathematical Information Processing, Space Research Institute of the Russian Academy of Sciences, 84/32 Profsoyuznaya Str, Moscow, 117997, Russian Federation*
[b]*Lavochkin Research and Production Association, 24 Leningradskoye Shosse, Khimki, 141400, Moscow Region, Russian Federation,*
[c]*Department of Dynamics and Flight Control of Rockets and Spacecrafts, Bauman Moscow State Technical University, 10 Gospitalnyi street, Moscow, 105005, Russian Federation*
[d]*Department of Probability Theory and Computer Modeling, Moscow Aviation Institute (National Research University), 4 Volokolamskoe shosse, Moscow, 125993, Russian Federation,*

**Abstract**

A problem of determining attainable landing sites on the surface of Venus is an essential part of the Venera-D project aimed to explore the planet using a lander. This problem appears due to the inability for the descent module to land at any point on the surface of Venus because of the short duration of the launch window (about 2 weeks from the optimal launch date), as well as restrictions on the maximum permissible overload. An additional factor affecting the reduction of attainable landing sites is the low angular velocity of Venus' own rotation. This study proposes a new approach to expand the attainable landing areas. The approach is based on the use of the gravitational field of Venus to transfer the spacecraft to an orbit resonant to the Venusian one with a ratio of periods of 1:1. All the simulations were performed at the patched conic approximation. As an example, we considered a flight to Venus at launch in 2029 or 2031. For both cases maps of attainable landing areas on the surface were constructed. It has been demonstrated that there is always at least one launch date within the launch window, allowing the spacecraft to reach almost any point on the surface of Venus. It is shown that the application of the proposed approach makes it possible to achieve a significant expansion of the attainable landing areas (over 70% of the surface) and, in some cases, provide access to any point on the surface of Venus. However, the price of this advantage is an increase in the flight duration by one Venusian year.

**Nomenclature**

$e$ = the eccentricity of the incoming hyperbolic orbit;
$p$ = the focal parameter of the incoming hyperbolic orbit, km;
$r_{p1}$ = the radius of periapsis for the landing trajectory, km;
$r_{p2}$ = the radius of periapsis of the flyby trajectory, km;
$r$ = the planetocentric distance at the point of spacecraft entry into the atmosphere, km;
$\Delta V$ = characteristic velocity, km/s;
$V$ = velocity of the spacecraft at the entry point, km/s;
$\mathbf{V}_\infty = \mathbf{V}_{r0}$ = vector of the hyperbolic excess velocity of the spacecraft, km/s;
$V_{r0} = |\mathbf{V}_\infty|$ = magnitude of $V_\infty$, km/s;
$\mathbf{V}_p$ = the Venus orbital velocity, km/s;
$\mathbf{V}_{a0}$ = the heliocentric velocity of the spacecraft before Venus flyby, km/s;

$\mathbf{V}_{ai}$ = the heliocentric velocities of the spacecraft after Venus flyby which corresponds to orbits with the same orbital period as the Venusian one, km/s;

$\mathbf{V}_{ri}$ = relative velocities of the spacecraft that can be reached after Venus gravity assist, km/s;

$α_{min}$ = the minimum turn angle required to obtain the relative velocity $\mathbf{V}_{r1}$, deg;

$α_{max}$ = the maximum turn angle required to obtain the relative velocity $\mathbf{V}_{r2}$, deg;

$α$ = the turn angle, i.e. angle between incoming and outgoing asymptotes of the spacecraft's hyperbolic approach trajectory on the pericenter height, deg;

$α*$ = the turn angle available by gravity field of the planet for chosen pericenter, deg;

$δ$ = the angle between $\mathbf{V}_p$ and $\mathbf{V}_{r0}$, deg;

$γ$ = the angle between the direction normal to the Venus orbit and the projection of the $\mathbf{V}_{rn}$ on the plane perpendicular to $\mathbf{V}_p$, deg;

$γ_0$ = initial value of $γ$ determined by vectors $\mathbf{V}_{r0}$ and $\mathbf{V}_p$ and by angles $α_{min}$ and $δ$, deg;

$θ$ = entry angle into Venus atmosphere, i.e. the angle between the spacecraft velocity and the local horizon at entry point, deg;

$ϑ$ = magnitude of the true anomaly of the spacecraft at the point of entry into the Venus atmosphere, deg.

$μ$ = 324859 $km^3/s^2$ is the gravitational parameter of the Venus.

$φ$ = angular radius of the circle of virtual pericenters, deg;

$ψ$ = angular radius of landing circle, deg;

## 1. Introduction

After astronomers discovered traces of phosphine in the atmosphere of Venus in 2020 [1], the hypothesis of the existence of life on Venus became widely discussed. It is assumed that life may exist not on the surface of the planet itself, but in the lower/middle cloud deck [1], since in this area the pressure is approximately equal to the Earth's pressure (1 atm), and clouds mainly consist of sulfuric acid, which can contribute to the development of microorganisms. In particular a number of space missions are being prepared in order to answer the most important question of the habitability of Venus. NASA is developing two such missions: DAVINCI+ [1] and VERITAS [2]. The purpose of the first mission is to study the atmosphere of Venus during two flybys of the planet, as well as to explore the proposed landing site (Alpha Regio is currently selected) and land after the second flyby. The purpose of the second mission is an observing the atmosphere by an orbiter during a long period of time, as well as conducting remote sensing of Venus.

One of the most prominent projects to explore the surface of Venus and its atmosphere using a lander and orbiter is the Russian Venera-D project [3] [2, 3, 4]. The first launch of the mission spacecraft is expected in 2029. The spacecraft for the mission will include a lander and an orbital module. The lander will also include a long-lived (lifetime about 1 month) and a short-lived (lifetime about a few hours) scientific stations [3, 4]. As an extended part of the Venera-D mission, a concept of sending a small spacecraft to the $L_1$ libration point of the Sun-Venus system to observe the Venusian atmosphere was considered [5].

An important part of a mission to explore Venus with landing stage is the selection of sites on the surface attainable for landing. The choice of a specific landing site for the Venus expedition should be determined by landing safety and scientific relevance criteria as well as constrained by the mission ballistics. Ballistics is the most important one, as it determines one of the most critical parameters of a space mission: the payload mass to be delivered to Venus. At the same time a number of other ballistics constraints may also be considered, such as, the limits on launch of the spacecraft from Earth only within a short time window, when the flight to Venus is the most advantageous in terms of the payload's mass, limitations on trajectories of the orbiter and the lander , and maximum permissible overload that the lander experiences during the descent in the dense Venusian atmosphere, etc. Thus, based on ballistic constraints, landing sites may be limited to a fairly small part of the surface of Venus (less than 5%).

---

[1]DAVINCI+ = Deep Atmosphere Venus Investigation of Noble gases, Chemistry, and Imaging, Plus is the NASA planned misison for Venus exploration. The official we page of the project on the nasa web-site https://www.nasa.gov/feature/goddard/2021/nasa-to-explore-divergent-fate-of-earth-s-mysterious-twin-with-goddard-s-davinci

[2]VERITAS = Venus Emissivity, Radio Science, InSAR, Topography, and Spectroscopy is the NASA planned misison for Venus exploration. The official web page: https://solarsystem.nasa.gov/missions/veritas/in-depth/

[3]Official web-page of the project: http://www.venera-d.cosmos.ru/index.php?id=658 (Accessed January 20, 2022)

The easiest way to increase the total number of attainable landing sites may be the enlarge of the launch window by moderate increasing the $\Delta V$ required for the flight to Venus. Results obtained in Ref. [6] show that such approach leads to growth in the launch $\Delta V$ at least on 200 m/s. Thus, this technique is very limited, primarily by the rapid increase in $\Delta V$ when the launch window is strongly extended.

As another approach, the spacecraft can be transferred into an orbit around Venus. In this way the total number of landing sites increases because the transfer to the landing trajectory may be performed from the orbit in the vicinity of the desired site compared to scenario where the landing is performed directly from the hyperbolic approach trajectory. Expansion of attainable landing areas in this case depends on the eccentricity of the spacecraft intermediate orbit. The higher eccentricities result in less attainable areas because the optimal descend may be performed only in the vicinity of periapsis. The best results are for circular orbit where descend may be performed anywhere along the orbit. However, the costs for those expansion is reversed. Spacecraft's break into a circular orbit around Venus requires at least about 3 km/s cost of $\Delta V$. Also, worth noting is the more complex design of the scenario itself, compared to a direct landing from an approach hyperbolic trajectory.

The attention should be given to the research on the possibility of expansion of the attainable landing areas using a maneuverable lander [7]. Usage of such vehicle allows to significantly increase the total number of landing sites. However, as such a vehicle has a special design, there are restrictions on the required entry angle into the atmosphere, because of maximum overload, as well as the requirements for a special descent trajectory into the atmosphere itself, which is accomplished by means of vehicle control. Thus, the complexity of mission increases.

In this research, it is proposed to transfer the spacecraft to an orbit resonant with Venus, with a ratio of the orbital periods of the spacecraft and the planet of 1:1 to expand attainable landing areas. For the transfer to the resonant orbit, it is supposed to use the gravity assist of Venus. In this case the trajectory parameters will be changed when the spacecraft returns to Venus after one revolution on the resonant orbit, thus it is possible to achieve a change of the landing site. There are many possible resonant orbits with ratio 1:1 of orbital periods, thus, there are many points on Venus surface attainable for landing. Notice, that the resonant orbits differs by eccentricity and inclination.

The proposed approach to expand the attainable landing areas depends on the entry angle and consequently on the maximum overload. In this paper the effect of entry angle on the number of attainable landing sites on the surface of Venus is studied. The study is made for entry angles of 7, 12 and 25 deg.

Note, that obtaining resonant orbits by gravity assist of celestial bodies in order to solve various problems of applied celestial mechanics has been widely known through the different groups of scientists. For example, one may highlight the results of Ref. [14] in which authors proposed to use resonant orbits for the spacecraft approach to Jupiter moons Ganymede and Callisto. Similar research was provided at [11] where authors developed flight schemes to Ganymede with gravity assists of Jovian's moons with the lowest radiation doze received during the flight. The other researchers from the Russian Keldysh Institute of Applied Mathematics developed ways to obtain orbits with high inclination to the ecliptic by a series of gravity assist of Venus [8, 9, 10]. It is worth noting that in research [12] the use of gravity assist maneuvers to obtain resonant orbits with Saturn moons is considered.

Finally, it should be noted that in previous studies we described the discussed concept in general and analyzed attainable landing areas for launch dates in 2029-2034 (for example, see Refs.[13, 20]). In this paper, however, we focus on a detailed description of the method, as well as on an algorithm for obtaining landing sites. The dynamics of the change in the attainable landing areas within the launch windows in 2029 and 2031 are also provided. The application of the method using the example of landing in the Imdr region is demonstrated.

## 2. Mathematical methods

### 2.1. Interplanetary trajectory calculation

In order to simplify the calculation and optimization of Earth-Venus trajectories, a method of patched conic approximation is used [15]. According to it the spacecraft trajectory is separated into planet-centric and heliocentric parts. Thus, the n-body problem is split into n two-body problems. Planets spheres of influence (SOI) in a heliocentric motion are reduced to a point, and at the planet-centric trajectory part SOI are considered to have infinite dimensions. Because of this, the spacecraft velocity at the boundary of the planet's SOI is assumed to be equal to the $V_\infty$. The spacecraft's trajectory within the planetary SOI is calculated according to the Keplerian theory. The heliocentric part of the trajectory, connecting with two planet-centric sections, is determined by the solution of the Lambert problem. The solution of the Lambert problem determinates the spacecraft trajectory by two given initial positions of the spacecraft, between which the flight is made, and the duration of such a flight. There are many methods to solve the Lambert problem, for example Ref. [16] describes the one that has been used by the authors.

The optimization of interplanetary trajectories is carried out by using the stochastic gradient descent algorithm, which uses the numerical approximation of Jacobian for finding the minima. The sum of launch $\Delta V$ ($\Delta V_0$) and impulse ($\Delta V_1$) required to insert satellite to the elliptical orbit around Venus was used as an optimization criterion. Note, that the analysis of the trajectory of the orbiter is not discussed in this article.

## 2.2. Design of the Near-Venus part of the trajectory

Let us analyse landing on the surface of Venus from a hyperbolic approach trajectory. When the spacecraft enters the Venusian SOI, a bunch of incoming hyperbolic trajectories is formed (Fig. 1), which forms a cylindrical surface with a diameter equal to the minor semi-axis of the incoming hyperbolic trajectory. Each of the trajectories shown in Fig. 1 is reachable for the spacecraft by means of small deviation of the spacecraft approach velocity vector at the boundary of Venusian SOI. Thus, any of these trajectories provides entry into the atmosphere of Venus at the required angle. Adopted in this work a value of entry angle equals to 25 degrees, according to [17] this value was used as a basic one for Soviet Venus exploration missions. Also, landing with entry angles of 7 and 12 degrees is considered below.

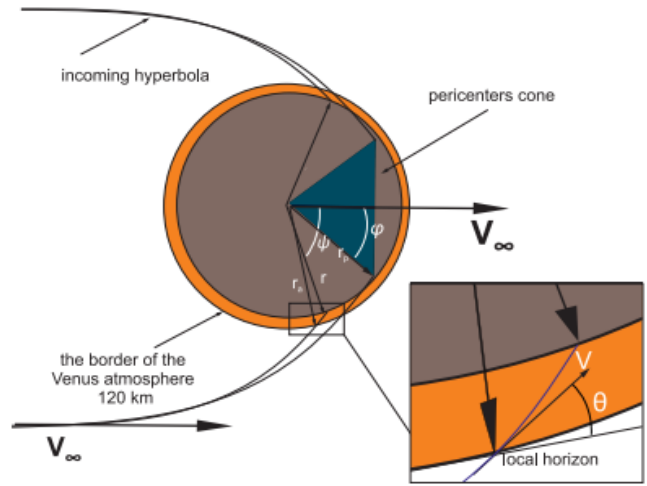

Figure 1: Geometry of the approaching part of the spacecraft trajectory.

The geometry of the approaching part of the trajectory is shown in Fig. 1. In the picture the hyperbolic excess velocity vector is shown passing through the center of Venus. The hyperbolic trajectories in Fig. 1 are going along the formations of a circular cylinder along this vector. Intersections of the hyperbolic trajectories with the surface of Venus form landing circles with angular radius $\psi$. The angle $\psi$ is calculated as follows (Ref. [18]):

$$\psi = \varphi + \vartheta \tag{1}$$

The angular radius of the circle of virtual pericenters and the magnitude of the true anomaly can be calculated as follows (Ref. [18]):

$$\cos\varphi = \frac{1}{1 + r_{p1} V_{r0}^2 / \mu} \tag{2}$$

$$\tan\theta = \frac{er\sin\vartheta}{p} \tag{3}$$

Note that the eq. (1)-(3) permit obtaining the parameters of the lander trajectory at the point of entry into dense layers of the atmosphere (at an altitude of about 120 km). The further part of the trajectory is integrated by the Runge-Kutta method of order 8(9) with an automatic control of the calculation error.

In the research a model of a spherical descent vehicle with a diameter of about 2.4 m and a mass of about two tons was assumed. The calculations were made using the model of Venusian atmosphere at an altitude below 140 km based on data obtained by previous Venuses exploration missions (see Ref. [19]).

## 3. Venus gravity assist for expanding attainable landing areas

General concept of the proposed method using gravity assist of Venus to transfer the spacecraft to a resonant orbit to provide landing to desirable region of the Venus surface is shown at Fig. 2. Overall, the main phases of the mission can be highlighted as follows: (I) Earth to Venus stage, (II) transfer to the required resonant orbit using a gravity assist of Venus, (III) waiting in the resonant orbit for 224.7 days, (IV) landing at the target site. In this section especial attention will be given to the second (II) stage.

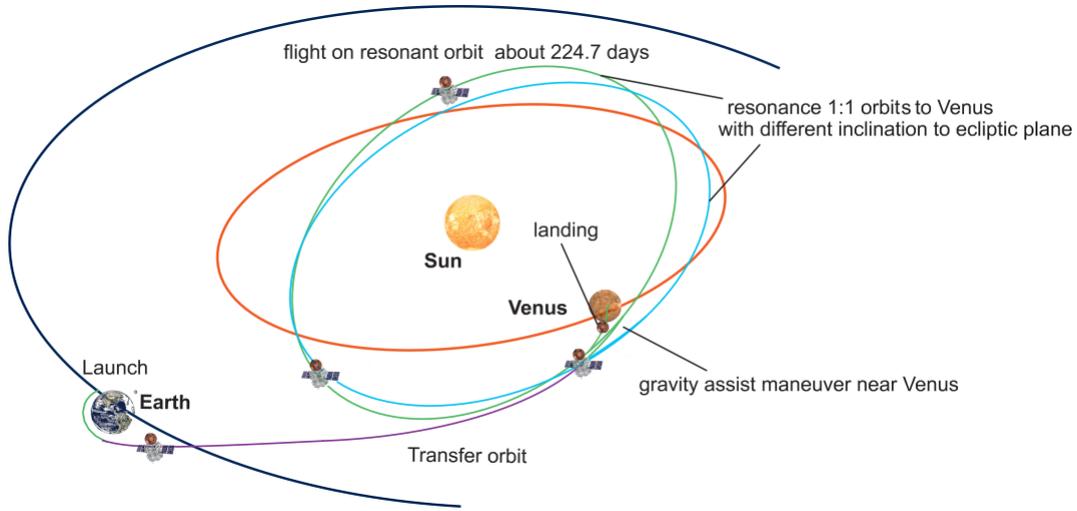

Figure 2: The proposed concept of using resonant orbits to expand landing areas.

The operations performed on stage (II) can be described in the Keplerian approximation by the following steps (Fig. 3):

1. A sphere with the radius equal to value of the Venus velocity vector $\mathbf{V}_p$ (in the heliocentric motion) and the center coinciding with the planet's mass center is formed.
2. A sphere with the radius equal to the value $V_{r0}$ represents all possible directions of turn of the incoming relative spacecraft velocity vector $\mathbf{V}_{r0}$ centered at the end of the $\mathbf{V}_p$.
3. The intersection of the above spheres is a circle, which is the base of two cones. These cones are formed respectively by a set of all possible vectors $\mathbf{V}_{ai}$ of heliocentric velocity of the spacecraft after Venus flyby (this cone is represented on Fig. 3 by vectors $\mathbf{V}_{a1}$, $\mathbf{V}_{a2}$ and $\mathbf{V}_{an}$) and a set of corresponding vectors $\mathbf{V}_{ri}$ of the relative spacecraft velocity (this cone is represented on Fig. 3 by vectors $\mathbf{V}_{r1}$, $\mathbf{V}_{r2}$ and $\mathbf{V}_{rn}$). Note that vectors $\mathbf{V}_{ai}$ belonging to the first cone are equal in magnitude to the vector of orbital velocity of Venus.
   Note that in this case:

$$\mathbf{V}_{ai} = \mathbf{V}_{ri} + \mathbf{V}_p, \, i = \overline{1, n} \qquad (4)$$

The above spheres and the circle of their intersection are shown in Fig. 3, which also shows the turn angle $\alpha^*$ which can be found from the equation:

$$\sin\frac{\alpha}{2} = \frac{1}{1 + r_{p2} V_{r0}^2 / \mu} \qquad (5)$$

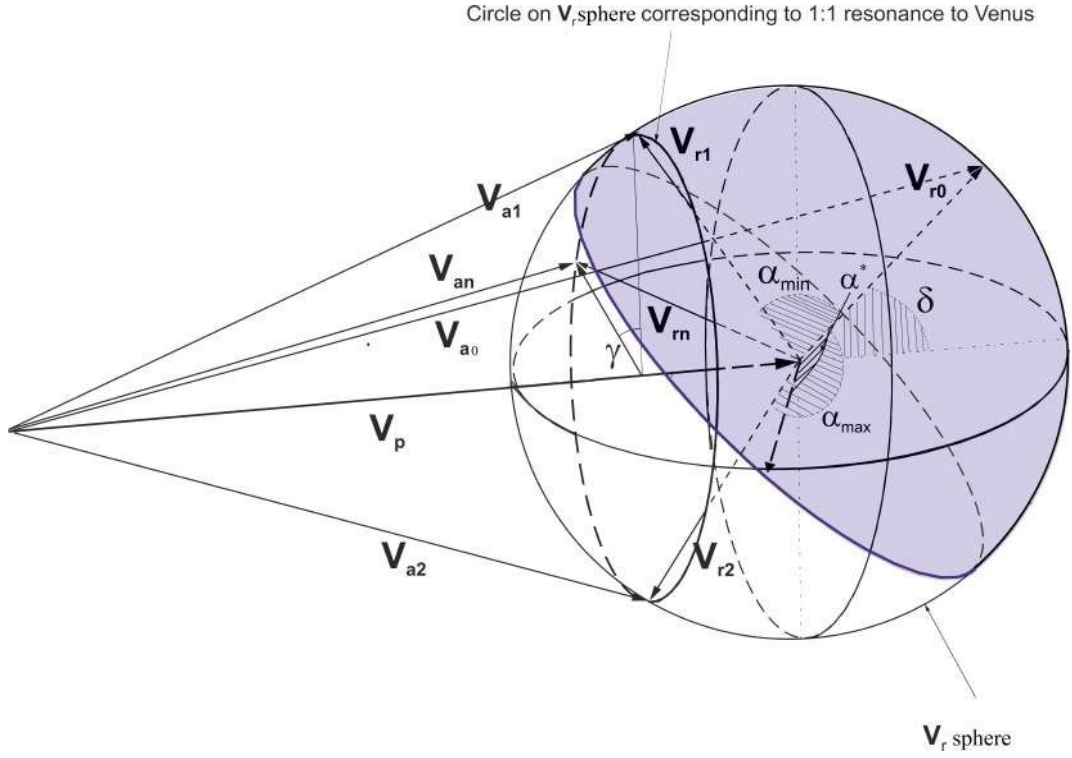

Figure 3: Geometry of a gravity assist maneuver with consequent spacecraft transfer to the orbit with resonance of 1:1 ratio to the Venusian one.

Note, that eq. (5) is essentially a version of previously given eq. (2). The angle $α^*$ is calculated for chosen radius of periapsis $r_{p2}$ and determinates whether turn of relative velocity from incoming to outgoing direction is available or not. The value of $r_{p2}$ is determined from an assumption that incoming and outgoing asymptotes are connected at the pericenter of the flyby trajectory and depends on the $α$ angle. Since the minimal value of the periapsis radii is equal to 6551 km this value was chosen for preventing the effect of the Venusian atmosphere on the spacecraft's flyby trajectory. The upper value of radius is not restricted.

4. The plane of rotation is defined as the plane formed by the vector of approach velocity $\mathbf{V}_{r0}$ and a velocity vector from the set of vectors $\mathbf{V}_{ri}$, defining a resonant orbit with the beginning at the end of a vector of velocity of Venus and the end on the mentioned circle, which ensures the getting of the vehicle to the set point on the surface of Venus during its return to the planet in a Venusian year.

5. $\mathbf{V}_{r0}$ at Venus is known from the solution of the Lambert problem. The required $\mathbf{V}_{ri}$ after the gravity assist maneuver is chosen by turning by the angle $α_{min}$ into the plane formed by $\mathbf{V}_{r0}$ and $\mathbf{V}_p$. The required $\mathbf{V}_{ri}$ lies on the surface of a cone, the base of which is the intersection circle of the two spheres. The $\mathbf{V}_{ri}$, providing landing at chosen point on the surface of Venus from the set belonging to the given cone, is reached by additional turn by the angle $γ$ counted clockwise in a positive direction from the Venus pole. The plane in which the angle $γ$ is defined is orthogonal to the direction $\mathbf{V}_p$ and coincides with the intersection circle of the two previously described velocity spheres (in other words it can be interpreted as the azimuth if the mentioned circle of possible resonant orbits would be projected onto surface of Venus). The angle $γ$ is a projected parameter depending on the choice of a landing point. A detailed algorithm for obtaining required $γ$ that corresponds to the desired landing point will be described below.

6. The condition of accessibility of all set of vectors whose ends lie on the circle of intersection of spheres (fig. 3) and, accordingly, provide equality of magnitudes of spacecraft's heliocentric velocities and Venusian heliocentric one, can be written in the form:

$$α_{max} \leq α^* \qquad (6)$$

In eq. (6) the value of the required $\alpha_{max}$ should be smaller than the value of $\alpha^*$ from eq. (5).

Let us describe the algorithm of obtaining the angle $\gamma$ which is a design parameter and depending the initial landing point chosen. That angle relates the desired landing point on the Venus surface and the vectors $\mathbf{V}_{ri}$ from the $V_r$ sphere required to transfer to the resonant orbit after Venus gravity assist.

Let us draw a circle with a center at the planned landing point ($L$) and a radius equal to the radii of the landing circles $\psi$ (Fig. 4). The resulting circle will intersect the circle of projections of possible velocities of the spacecraft after the gravity assist maneuver in two points (A and B). The points of intersection are the projections of the required spacecraft velocity vectors, i.e. in the general case, the problem will have two solutions. The angles $\gamma_A$ and $\gamma_B$ corresponding to the obtained points are the desired angles, which are defined as follows [21]:

$$\gamma_{(A,B)} = \gamma_C \pm \arccos \frac{\cos \psi}{\cos |\xi - \sigma|} \quad (7)$$

where $\gamma_C$ is the angle between the vertical axis of the chosen coordinate system and the radius-vector of the landing point (Fig. 4); $\sigma$ is the value of the arc $PL$, which can be found by the coordinates of points $P$ and $L$ through the solution of the inverse geodetic problem and $\xi$ is the value of the arc $PC$.

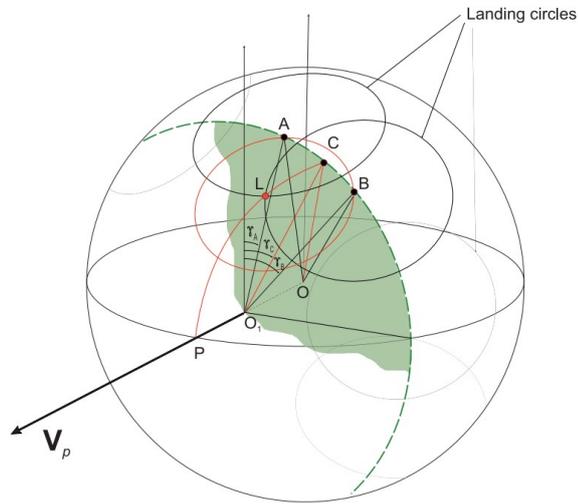

Figure 4: The geometry of determination of $\gamma$ angle. The sphere corresponds to the Venus surface.

Two special cases might be considered:

1. $\Delta\gamma = 0$. In this case the given landing point lies on the boundary of the tolerance area. In this case $\gamma_A = \gamma_B = \gamma_C$.
2. $|\xi - \sigma| = 0$. In this case, the given landing point lies on the circumference of the velocity projection of the spacecraft after the gravity assist maneuver. In this case $\gamma_{A,B} = \gamma_C \pm \psi$.

## 4. Influence of an entry angle on size of attainable landing areas

Note that the proposed approach to expand the attainable landing areas depends on the magnitude of the maximum overload suffered by the lander during the landing phase. As the maximum overload increases, the radius of attainable landing areas will increase, and vice versa as the maximum overload decreases. Since the maximum overload is directly proportional to the entry angle, the radius of the landing circle will also be related to the entry angle. This relationship can be represented by a simple formula and can be written as follows

$$\psi(\theta) = \arccos \frac{1}{1 + r_{p1}(\theta)V_{r0}^2/\mu} + \arcsin \frac{p \tan \theta}{er} \quad (8)$$

The equation (8) was obtained using (1), taking into account (2) and (3), at the following values of parameters r = 6,171 km, $V_{r0}$ = 3 km/s, $\mu$ = 324,859 $km^3/s^2$.

Dependence of the angular radius of landing circle vs. entry angle into Venusian atmosphere is shown in Fig. 5.

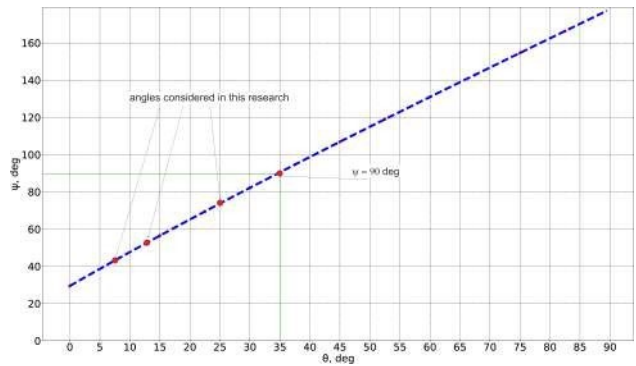

Figure 5: Radius of landing circle vs. entry angle into Venusian atmosphere.

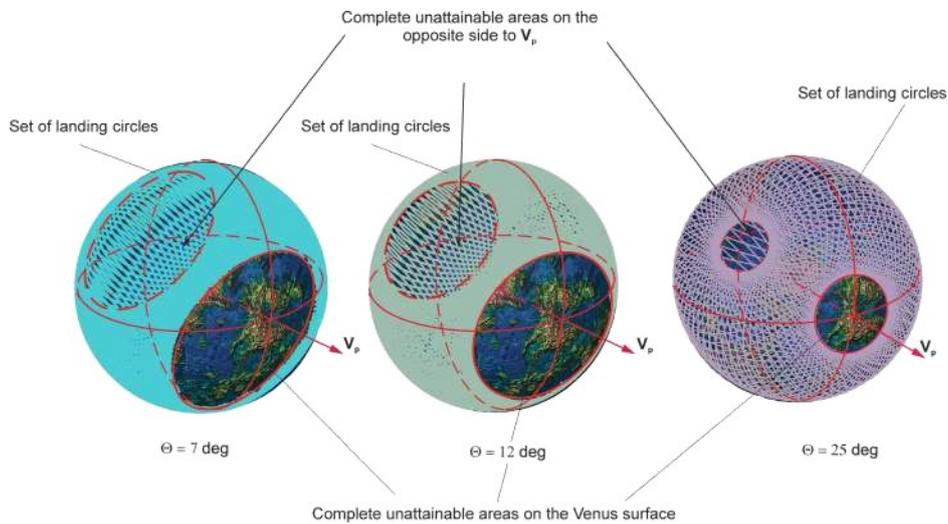

Figure 6: Unattainable landing areas for entry angles of 7, 12 and 25 deg.

Fig. 5 shows that due to the chosen entry angle, the magnitude of the angular radius is less than 90 degrees. Because of this, there are limitations on the attainable landing areas. Note that 90 degrees is achievable at entry angle of about 34 degrees. In our case, however, for entry angles of 7, 12 and 25 degrees the unattainable areas will cover 50 down to 15 degrees. These areas, for the entry angles considered are presented in Fig. 6.

## 5. Results

### 5.1. Landing at the desired place on the Venus surface at launch in 2029 using gravity assist

To show efficiency of discussed approach let us give some example of its application. To do that the region on the Venus surface the flight to which is almost impossible by direct flight might be chosen. Such a region may be chosen in the vicinity of young volcano formation sites, e.g. wide volcano Idunn Mons (at $46^oS$, $145^oW$) (Fig. 7). Landing of the descent vehicle in this region could be of interest as in such an area there are active geological and atmospheric processes (associated with intensive emission of volatiles into the atmosphere of Venus by volcanoes)[22]. Note that previously this region was studied only from the satellite orbit, landers did not reach Imdr region.

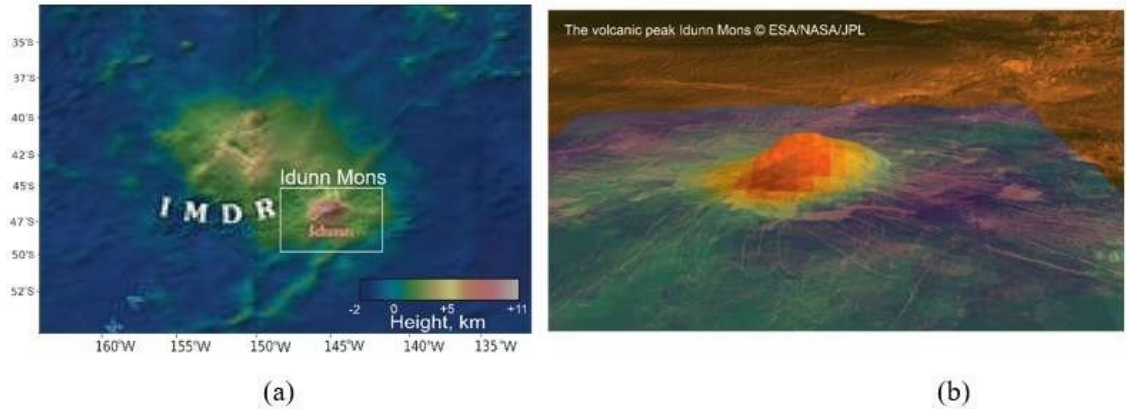

Figure 7: Imdr Regio (a); the volcanic peak Idunn Mons (at $46^{\circ}S$, $145^{\circ}W$) in the Imdr Regio area of Venus (© ESA/NASA/JPL) (b).

Let us consider landing in the given region at launch in 02.11-23.11.2029. Compare two approaches of landing: the commonly used strategy (direct landing from incoming hyperbola) (Fig 8a) and the proposed approach (Fig 8b). For modelling the final stage of landing, i.e. descent in dense layers of Venus atmosphere let us accept the entry angle value is 12 deg, the height of dense layer is 120 km and the ballistic coefficient is 0.0050 $m^2/kg$. Simulation of the final stage from entering into dense layers of Venusian atmosphere to arriving at the landing point was performed using Runge-Kutta method of 8(9) order.

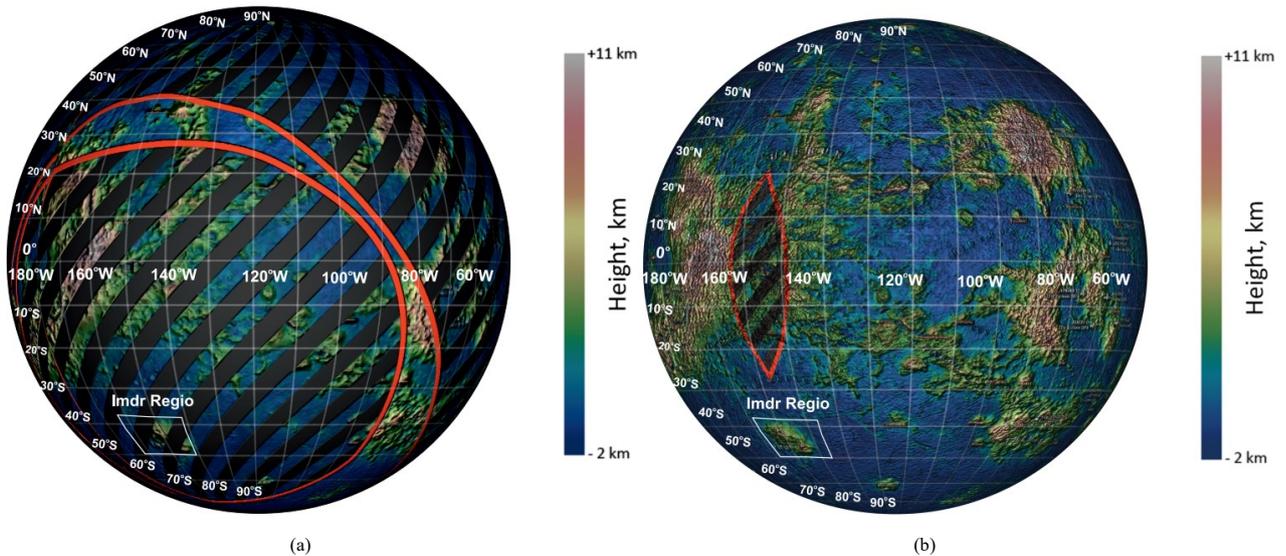

Figure 8: Unattainable landing areas for launch window within 02.11.2029-23.11.2029 (a) for landing from direct flight (b) using gravity assist.

As it can be seen from Fig. 8 the Imdr Regio cannot be reached by a direct flight to Venus. However, using the method described above, it is possible to reach Imdr Regio launching at any date in he considered launch window (Fig. 8b).

*5.2. Results for launch windows in 2029 and 2031*

As a part of this study, launch dates for the flight to Venus in 2029 and 2031 were considered. Fig. 9a shows the launch ΔV vs. launch and arrival dates to Venus. Trajectories to Venus are separated into trajectories of the first semi-turn, if the transfer angle is less than 180 degrees, and trajectories of the second semi-turn, if the transfer angleis more than 180 degrees.

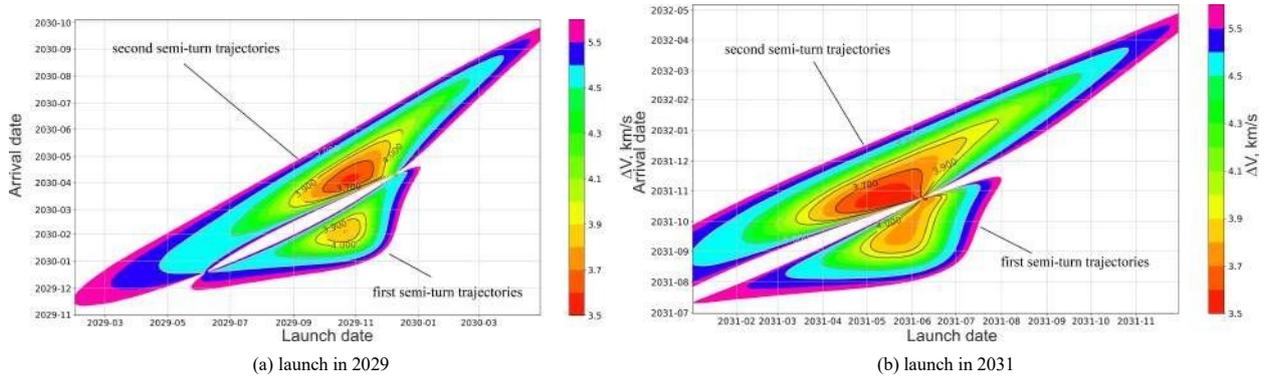

(a) launch in 2029   (b) launch in 2031

Figure 9: ΔV required for the flight to Venus vs the launch date and arrival date.

As can be seen from the figures, the ΔV costs for the second semi-turn trajectories at most launch date intervals are lower than those for the first semi-turn trajectories. But in order to reduce time of flight for the whole mission below we will consider only first semi-turn trajectories. Next, we will consider launch dates within the 2029 and 2031 launch windows. As these windows we will take the launch dates intervals 02.11.2029-03.12.2029, 20.05.2031-24.06.2031, within a month from the optimal launch date. Some of the spacecraft trajectory characteristics for launch in these dates are given in Appendix A.

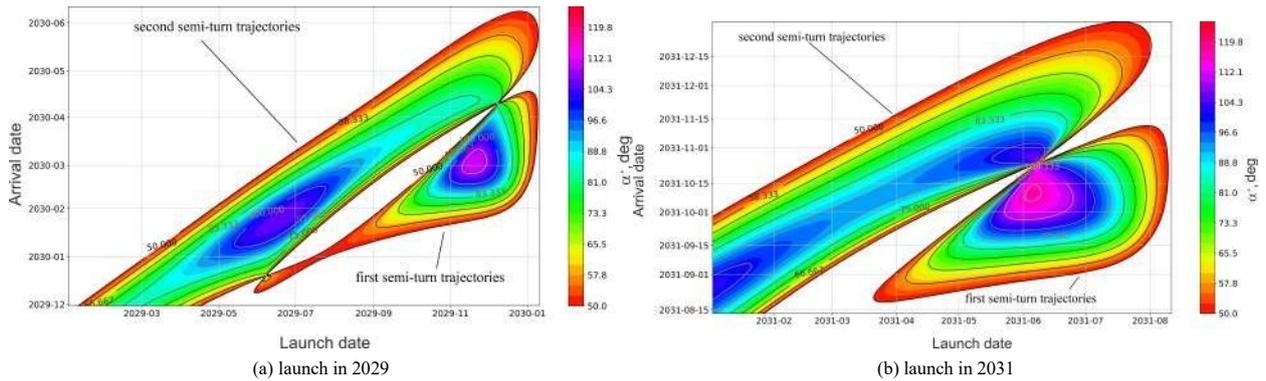

(a) launch in 2029   (b) launch in 2031

Figure 10: $\alpha*$ vs. launch date and arrival date.

Fig. 10a - 10b shows the dependence of the value of angle $\alpha*$ on the launch and arrival dates for spacecraft launches during launch windows in 2029 and 2031 to the launch date that provides a minimum ΔV for the flight along the first semi-turn trajectories (Fig. 9a-9b).

Let us plot the attainable landing areas on the map of Venus for different values of the entry angle. The results obtained for entry angle values of 7, 12 and 25 degrees are shown in Figs. 11-13. Note that these figures show cases where the required turn angle (eq. (5)) is achievable for the entire circle (i.e. ineq. (6) is satisfied). In this case the size of unattainable area varies from 50 down to 15 degrees depending on the accepted value of the entry angle (7, 12, 25 degrees).

In Figs. 10a-10b the unattainable landing areas are shown in dark color, each one corresponds to one launch date in 2029 and 2031. The center of those areas is the virtual point where vector of the heliocentric velocity of Venus intersects with planet's surface. Since the position of the unattainable areas varies for different launch-landing dates, this means that it is possible to change the landing site by changing the date of arrival to Venus. However, such a change is limited by the rapidly increasing characteristic velocity required for a launch from Earth.

The size of the area in which the outlined approach to eliminate unattainable areas can be applied is limited by the intersection of the curves that specify the size of the unattainable landing areas for the two dates: the start and end

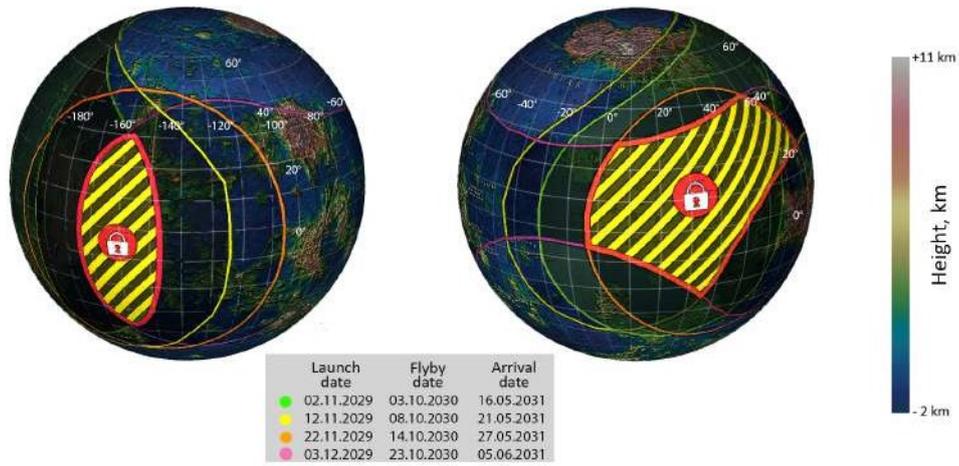

(a) launch in 2029

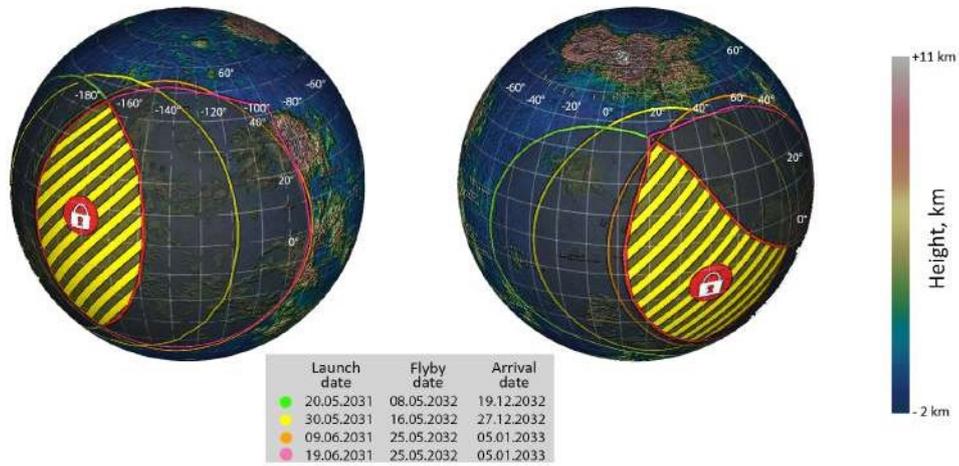

(b) launch in 2031

Figure 11: Unattainable landing areas (darkened) for launch windows in 2029 (a) and 2031 (b) using Venus gravity assist; the entry angle of the final stage of landing is 7 deg. The "locked" areas show the complete unattainable areas for landing.

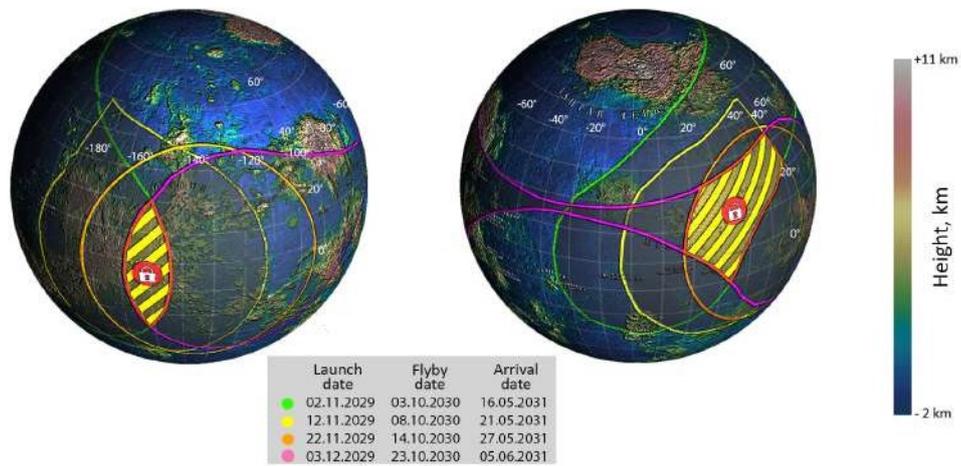

(a) launch in 2029

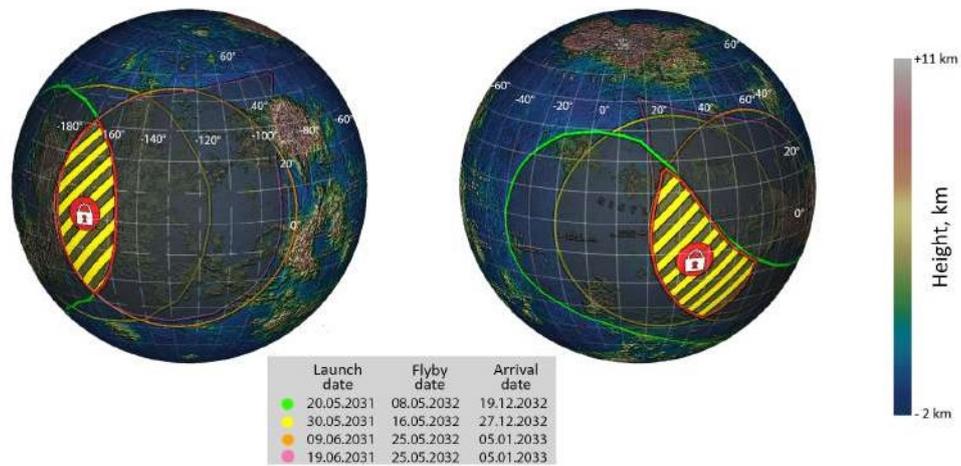

(b) launch in 2031

Figure 12: Unattainable landing areas (darkened) for launch windows in 2029 (a) and 2031 (b) using Venus gravity assist; the entry angle of the final stage of landing is 12 deg. The "locked" areas show the complete unattainable areas for landing.

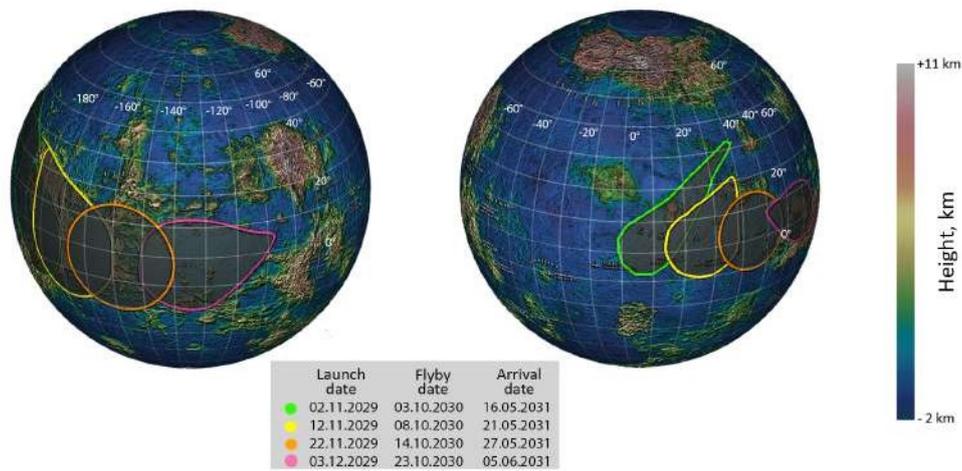

(a) launch in 2029

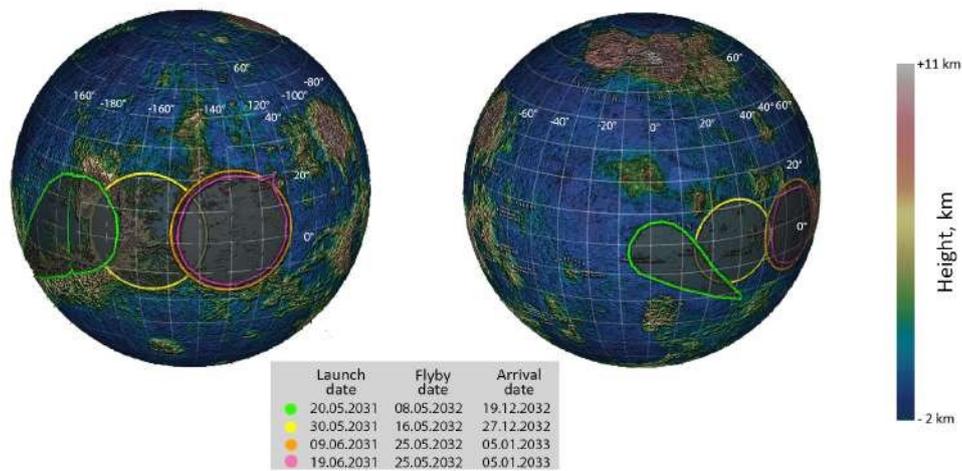

(b) launch in 2031

Figure 13: Unattainable landing areas (darkened) for launch windows in 2029 (a) and 2031 (b) using Venus gravity assist; the entry angle of the final stage of landing is 25 deg.

ones for the launch window shows an area ('locked' area in Fig. 10a-10b) that is completely unattainable for landing regardless of the launch date within the window. Notice, the size of the completely unattainable area will decrease as the entry angle increases (see sec. 4).

It is worth noting that in most of the considered cases, the shapes of the curves outlining the dark regions of unattainable points in Figs. 11-13 are close to circular with radii from about 50 deg (for the entry angle of 7 deg) to 15 deg (for an entry angle of 25 deg). However, the shapes of the boundaries of some areas are far from circular. This is particularly noticeable in Fig. 13, where some of the unattainable regions have a shape, which we will hereafter refer to as circle with a gap. The gap appears because ineq. (6) is violated in these cases. This means that the transfer to a part of the 1:1 ratio resonant orbits is not available to the spacecraft by Venus gravity assist because of this some approach trajectories are not feasible, resulting in a reduction in the number of landing circles and consequently making some areas on the surface of Venus unattainable.

### 5.3. Avoiding unattainable areas for the fixed launch date

As we mentioned above, unattainable areas shown in Fig. 14 appear due to the inability to rotate the $\mathbf{V}_{r0}$ by more than $\alpha_{max}$. Landing in one of those regions can be accomplished by varying the time of the first flyby of Venus. By varying the first flyby date on 7 days, it is possible to make the entire surface of Venus attainable for landing. Such an approach will be effective because Venus rotates about 1.5 degrees and about 1.6 degrees along the orbit per one day, meaning unattainable areas will move on the Venus surface. Thus, the previously unattainable areas will become attainable. The estimated cost of an additional $\Delta V_0$ of such flyby date variation is shown in Fig. 15.

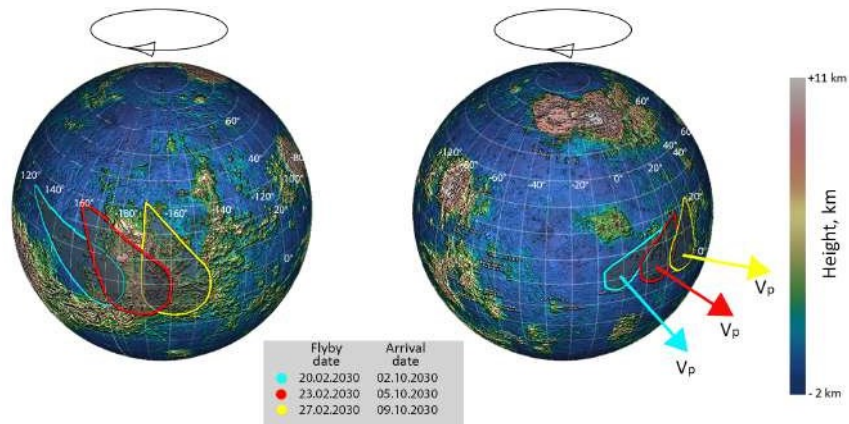

Figure 14: Areas of inaccessibility (shown in dark colour) for landing date 07.11.2029 with landing in 2030 through a revolution after a gravity assist maneuver in case not achieving all possible 1:1 resonant orbits with respect to Venus as a result of a gravity assist maneuver.

Note that the discussed approach making the whole surface of Venus attainable for landing may be applied only for entry angles of about 25 degrees and higher. Noteworthy, the shape and size of the unattainable area (circle or circle with gape as in Fig. 14) and that location on the Venus surface also affects the required range of dates in which flyby date-changing can be performed. For example, the above range of ±5 days is acceptable for the size of unattainable landing zones obtained for 25 deg value of entry angle and when they have the shape of a circle with a gap.

### 5.4. Note on the conditions of the Venus flyby

This subsection discusses the peculiarities of the Venus flyby for the transfer to the required resonant orbit, particularly the achievable pericenter radii and turn angles at which a transfer to a resonant orbit with Venus is possible.

Let us plot the radius of the pericenter of the flyby trajectory $r_{p2}$ and angle $\alpha$ vs. $\gamma$ for the launch dates 08.11.2029 and 08.06.2031.

Fig. 16 illustrates that for the launch date in 2029 the radius of periapsis changes in a wide range from 34,000 down to 6,551 km (the lower limit), while for 2031 the small change of $r_{p2}$ value in the range from 11,900 to 18,000 km is typical, which is generally associated with more favorable arrival conditions in 2031, as the velocity of Venus flyby is 2.93 km/s vs. 3.49 km/s in 2029. At the same time Fig. 16a shows that there is an area where the constraint (6) is violated, which is associated with reaching the minimum height of the pericenter (500 km). On the maps for the launch in 2029 (Fig. 11a, 12a, 13a) such a gap results in additional unattainable zones.

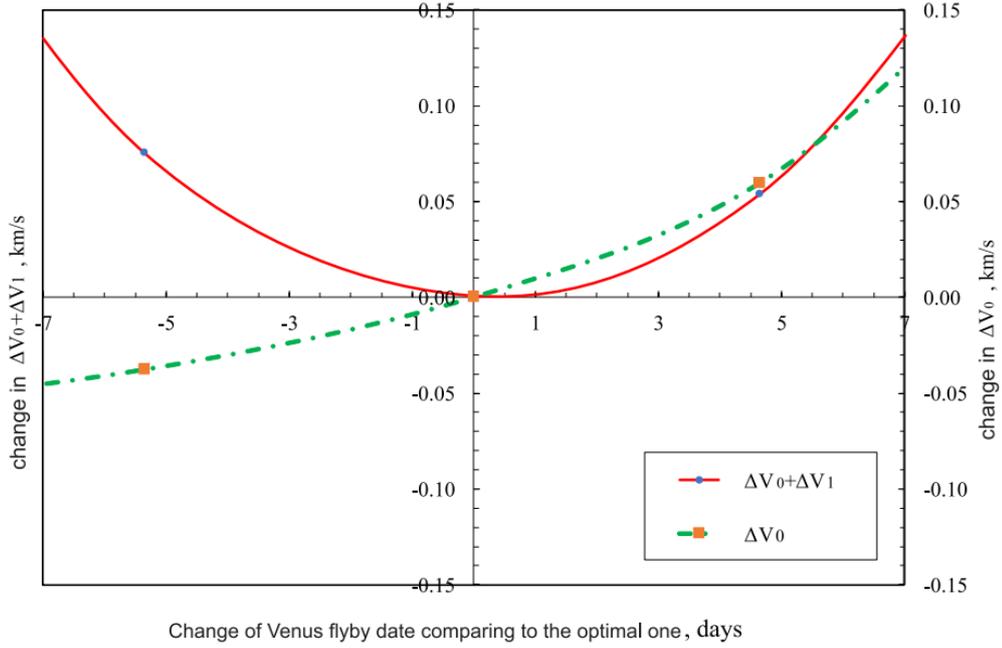

Figure 15: Change in $\Delta V_0$ and change in sum of impulses $\Delta V_0+\Delta V_1$ for launch on 07.11.2029 vs change of date of Venus flyby relative to its optimal values.

## 6. Discussion and Conclusions

The current research considers the usage of the Venus gravity assist in transferring the spacecraft to a 1:1 resonant orbit to Venusian one in order to expand the attainable landing areas on the Venus surface. The parameters of landing on the surface of Venus, namely the entry angle, which was supposed to be 7, 12, and 25 degrees, respectively, were used as constraints.

The very concept of using resonant orbits in the applications of celestial mechanics is known for a long time, for example, Beckman 1975 [23] and Uphoff 1976 [14] describe the use of gravity assist maneuvers to make $N\pi$ flights (i.e. with a whole number of revolutions) around the Jovian moons, and the use of their gravity to change the orbit inclination of the vehicle. Such a maneuvers are easy to calculate using a concept of $V_r$ sphere (V-infinity globe, $V_\infty$ sphere) (see sec. 3) [12, 9, 11].

Let us dwell on a series of works [11, 8, 9, 10] on the development of chains of gravity assist of Venus to obtain an orbit with a high inclination to the ecliptic in order to obtain an overview of the polar regions of the Sun. In a sense, the methods used by the authors are generally similar to those used in this paper, but the final objectives are very different. For example, to obtain a high inclination, the authors of [8, 10] needed to maximize the arrival asymptotic velocity, due to the Labunsky approximation [24], thereby sacrificing the maximum achievable turn angle $\alpha_{max}$. For example, to achieve an inclination to heliocentric ecliptic plane of 45 deg, the $\alpha_{max}$ will be 9.02 deg [9], at this angle $V_{r0} > 10$ km/s (under ineq. 6). However, in our problem, the key was the maximization of attainable landing areas, for which at first, it is required to increase the $\alpha_{max}$, which can be achieved by reducing the minimum $r_p$ to 6551 km, and more by reducing the $V_{r0}$. The latter was achieved by choosing the criterion for finding optimal trajectories: $\Delta V_0 + \Delta V_1$. Here it should be noted that, in general, this approach can be simplified to use the more classical criterion $\Delta V_0$, sincethe expansion of attainable landing areas will still be significantly larger than for a two-week launch window or even for a month extended case. However, the maximum advantage within the discussed technique is achieved exactly by choosing the criterion $\Delta V_0 + \Delta V_1$, which was generally shown in this work.

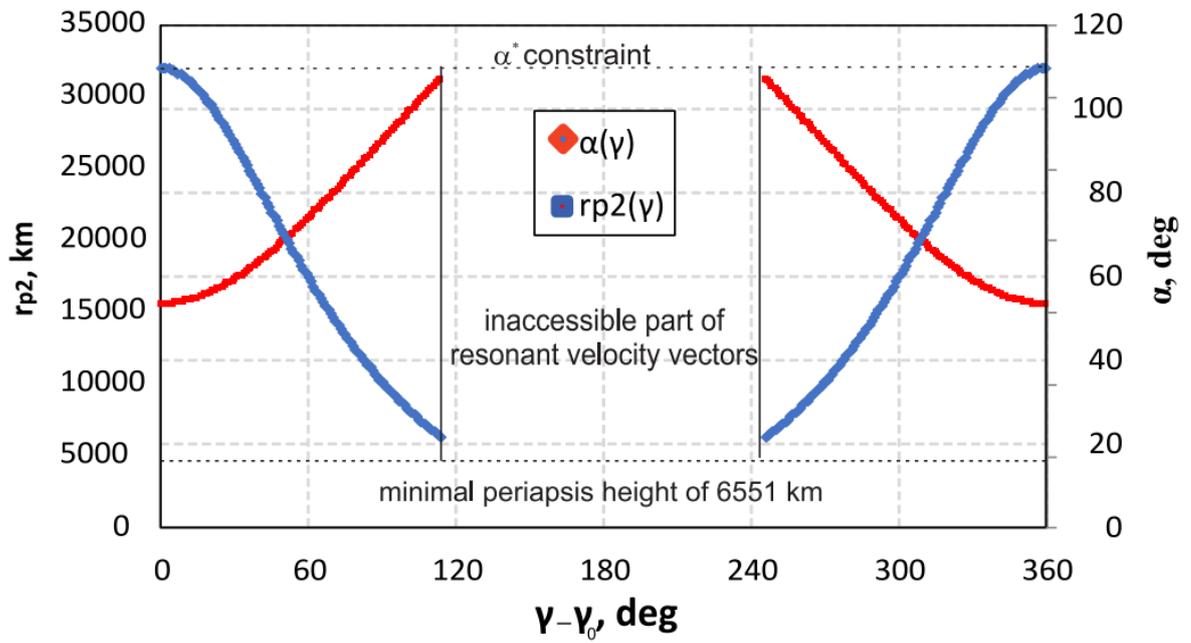

(a) launch in 08.06.2031

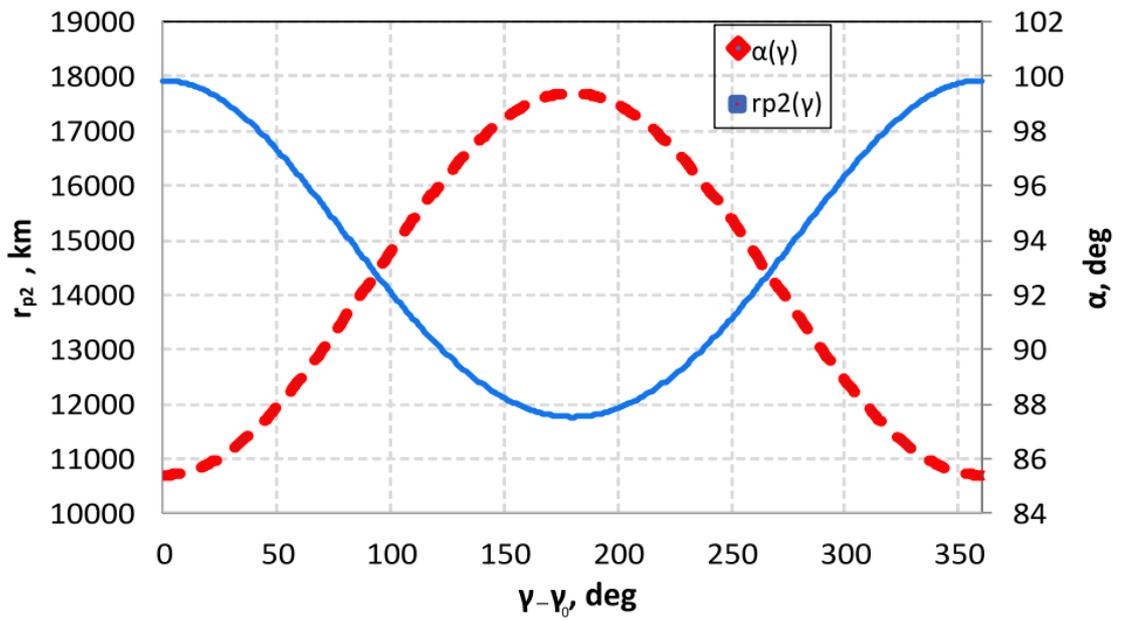

(b) launch in 08.06.2031

Figure 16: The dependence of $r_{p2}$ and the angle $α$ vs. the angle $γ$ for the launch dates: (a) - 06.11.2029; (b) - 08.06.201.

It is important to say a few words about currently developed NASA's project DAVINCI+ to study Venusian atmosphere and its surface. In the project scenario [25] the spacecraft is supposed to make two flyby of Venus in order to explore the surface of Venus to clarify the landing area, and then after the second flyby in 7 months to carry out a landing in the selected area on the surface [4]. Especially interesting for us is the concept of two gravity maneuvers near Venus, in all likelihood their purpose in addition to the mentioned above also serves to select the landing site, since the chosen area Alpha Regio is not attainable for landing from the direct flight path (without the gravity maneuver), that can be seen in [20]. But it is difficult to clarify or reject this assumption because no detailed scenarios with ballistics schemes of flight in framework of developing mission has been presented as of January, 2022.

Let us also discuss additional costs of the characteristic velocity required for correction maneuvers on the flight trajectory to Venus with the gravity assist maneuver. Indeed, since the Venus flyby is required to be performed with predetermined parameters, and due to the impossibility of maintaining absolute accuracy of their definition, a number of correction maneuvers will be required both during the first Venus flyby, as well as when returning to Venus. It should be noted here that the authors' calculations in the General Mission Analysis Tool (GMAT) [26] show that the value of such correction maneuvers will not exceed about 10-20% of total $\Delta V$. If we refer to the experience of the BepiColombo [27, 28], InterHeliozond [29, 30] (under development), and other missions, with multiply gravity assists, the additional correction maneuvers for our mission concept with one gravity assist will be significantly lower. However, it should be emphasized that in our simulation using a model including gravitational attraction of all the Solar System planets, oblateness of Earth and Venus, as well as solar radiation pressure the date of Venus flyby differs from the one obtained using the method of patched conic approximation, but in general such shift for this problem is of the order of a few hours.

Using the approach discussed in this paper, the attainable landing areas were calculated for launch in 2029 and 2031. As it is showed, the areas attainable for landing depend on the value of the entry angle. Thus, we studied the effect of the entry angle value of 7, 12, 25 deg on the size of the attainable landing areas. It is possible to state that for a value of 25 deg the whole surface of Venus becomes attainable for landing (Fig. 11- 13) except for two areas with an angular size of approximately 15 degrees. For making the whole surface to be attainable, it is needed to shift the dates of the first Venus gravity assist by no more than 10 days in total (Fig. 14). But when it comes to lower values of 7 and 12 deg the attainable landing areas decrease compared to 25 deg. On the other hand, the approach used in this article permits expanding landing areas for every launch date, even for small values of entry angle and makes at least half of Venus surface attainable for landing. In that way, the proposed approach provides a radical expansion compared to a simple direct flight to Venus.

---

[4]Currently as the landing site the Alpha Regio is chosen

## Appendix A. Parameters of the Venus approaching trajectories

Characteristics of the calculated trajectories required for obtaining attainable landing areas are given in tables A.1 and A.2. These tables show that the angles $α_{max}$ remain less than $α*$ for almost the entire duration of the launch dates intervals: from 02.11.2029 to 23.11.2029, from 26.05.2031 to 07.06.2031.

Table A.1: Characteristics of the spacecraft trajectory during the transit of Venus at launch in 2029 (trajectories of the first semi-turn)

| Launch date | Transit date | Landing date | $α_{min}$, deg | $α_{max}$, deg | $α*$, deg | $δ$, deg | $γ_0$, deg | $ΔV$, km/s | $V_∞$, km/s |
|---|---|---|---|---|---|---|---|---|---|
| 02.11.2029 | 20.02.2030 | 02.10.2030 | 46.1 | 140.1 | 101.5 | 47.0 | 136.9 | 3.93 | 3.80 |
| 03.11.2029 | 21.02.2030 | 03.10.2030 | 47.4 | 138.7 | 102.4 | 45.7 | 137.6 | 3.94 | 3.74 |
| 04.11.2029 | 21.02.2030 | 03.10.2030 | 48.7 | 137.3 | 103.3 | 44.3 | 138.3 | 3.95 | 3.69 |
| 05.11.2029 | 22.02.2030 | 04.10.2030 | 50.0 | 135.9 | 104.2 | 43.0 | 139.2 | 3.96 | 3.63 |
| 06.11.2029 | 23.02.2030 | 05.10.2030 | 51.4 | 134.5 | 105.1 | 41.5 | 140.1 | 3.97 | 3.59 |
| 07.11.2029 | 23.02.2030 | 05.10.2030 | 52.8 | 133.0 | 105.9 | 40.1 | 141.1 | 3.99 | 3.54 |
| 08.11.2029 | 24.02.2030 | 06.10.2030 | 54.3 | 131.4 | 106.7 | 38.6 | 142.1 | 4.01 | 3.49 |
| 09.11.2029 | 24.02.2030 | 06.10.2030 | 55.8 | 129.9 | 107.4 | 37.0 | 143.3 | 4.02 | 3.45 |
| 10.11.2029 | 25.02.2030 | 07.10.2030 | 57.3 | 128.2 | 108.1 | 35.5 | 144.6 | 4.03 | 3.41 |
| 11.11.2029 | 25.02.2030 | 07.10.2030 | 58.9 | 126.6 | 108.7 | 33.8 | 146.0 | 4.05 | 3.37 |
| 12.11.2029 | 26.02.2030 | 08.10.2030 | 60.5 | 124.9 | 109.3 | 32.2 | 147.6 | 4.07 | 3.34 |
| 13.11.2029 | 27.02.2030 | 09.10.2030 | 62.1 | 123.3 | 109.9 | 30.6 | 149.4 | 4.08 | 3.31 |
| 14.11.2029 | 27.02.2030 | 09.10.2030 | 63.8 | 121.6 | 110.3 | 28.9 | 151.4 | 4.11 | 3.29 |
| 15.11.2029 | 28.02.2030 | 10.10.2030 | 65.4 | 119.9 | 110.7 | 27.2 | 153.7 | 4.13 | 3.26 |
| 16.11.2029 | 28.02.2030 | 10.10.2030 | 67.1 | 118.2 | 111.0 | 25.6 | 156.2 | 4.15 | 3.25 |
| 17.11.2029 | 01.03.2030 | 11.10.2030 | 68.7 | 116.6 | 111.3 | 24.0 | 159.2 | 4.17 | 3.23 |
| 18.11.2029 | 02.03.2030 | 12.10.2030 | 70.2 | 115.1 | 111.4 | 22.4 | 166.4 | 4.20 | 3.22 |
| 19.11.2029 | 02.03.2030 | 12.10.2030 | 71.7 | 113.6 | 111.5 | 20.9 | 170.9 | 4.23 | 3.22 |
| 20.11.2029 | 03.03.2030 | 13.10.2030 | 73.1 | 112.2 | 111.5 | 19.6 | 176.1 | 4.26 | 3.22 |
| 21.11.2029 | 03.03.2030 | 13.10.2030 | 74.3 | 111.0 | 111.4 | 18.3 | 182.0 | 4.29 | 3.22 |
| 22.11.2029 | 04.03.2030 | 14.10.2030 | 75.4 | 109.9 | 111.3 | 17.3 | 188.7 | 4.32 | 3.23 |
| 23.11.2029 | 04.03.2030 | 14.10.2030 | 76.2 | 109.1 | 111.0 | 16.5 | 196.2 | 4.35 | 3.25 |

Table A.2: Characteristics of the spacecraft trajectory during the transit of Venus at launch in 2031 (trajectories of the first semi-turn)

| Launch date | Transit date | Landing date | $\alpha_{min}$, deg | $\alpha_{max}$, deg | $\alpha^*$, deg | $\delta$, deg | $\gamma_0$, deg | $\Delta V$, km/s | $V_\infty$, km/s |
|---|---|---|---|---|---|---|---|---|---|
| 26.05.2031 | 01.10.2031 | 12.05.2032 | 60.5 | 124.9 | 111.0 | 32.2 | 49.2 | 3.79 | 3.25 |
| 27.05.2031 | 02.10.2031 | 13.05.2032 | 62.7 | 122.6 | 112.0 | 29.9 | 48.3 | 3.79 | 3.19 |
| 28.05.2031 | 02.10.2031 | 13.05.2032 | 65.0 | 120.1 | 113.0 | 27.6 | 47.3 | 3.79 | 3.14 |
| 29.05.2031 | 03.10.2031 | 14.05.2032 | 67.4 | 117.6 | 113.9 | 25.1 | 46.0 | 3.79 | 3.09 |
| 30.05.2031 | 04.10.2031 | 15.05.2032 | 70.0 | 115.0 | 114.8 | 22.5 | 44.4 | 3.78 | 3.04 |
| 31.05.2031 | 05.10.2031 | 16.05.2032 | 72.6 | 112.3 | 115.5 | 19.9 | 42.4 | 3.78 | 3.00 |
| 01.06.2031 | 06.10.2031 | 17.05.2032 | 75.3 | 109.5 | 116.2 | 17.1 | 39.8 | 3.78 | 2.97 |
| 02.06.2031 | 07.10.2031 | 18.05.2032 | 78.1 | 106.7 | 116.7 | 14.3 | 36.3 | 3.78 | 2.94 |
| 03.06.2031 | 08.10.2031 | 19.05.2032 | 80.9 | 103.8 | 117.1 | 11.4 | 30.9 | 3.78 | 2.92 |
| 04.06.2031 | 09.10.2031 | 20.05.2032 | 83.7 | 101.0 | 117.4 | 8.7 | 22.2 | 3.78 | 2.90 |
| 05.06.2031 | 09.10.2031 | 20.05.2032 | 86.2 | 98.5 | 117.5 | 6.1 | 5.8 | 3.79 | 2.90 |
| 06.06.2031 | 10.10.2031 | 21.05.2032 | 87.8 | 96.9 | 117.5 | 4.6 | 333.8 | 3.79 | 2.90 |
| 07.06.2031 | 11.10.2031 | 22.05.2032 | 87.4 | 97.4 | 117.3 | 5.0 | 293.4 | 3.79 | 2.91 |
| 08.06.2031 | 12.10.2031 | 23.05.2032 | 85.4 | 99.4 | 116.9 | 7.0 | 268.3 | 3.79 | 2.93 |
| 09.06.2031 | 13.10.2031 | 24.05.2032 | 83.0 | 101.8 | 116.4 | 9.4 | 254.9 | 3.80 | 2.95 |
| 10.06.2031 | 13.10.2031 | 24.05.2032 | 80.8 | 104.1 | 115.9 | 11.6 | 246.4 | 3.80 | 2.98 |
| 11.06.2031 | 14.10.2031 | 25.05.2032 | 79.1 | 105.8 | 115.3 | 13.4 | 240.1 | 3.81 | 3.02 |
| 12.06.2031 | 14.10.2031 | 25.05.2032 | 77.9 | 107.1 | 114.8 | 14.6 | 235.0 | 3.83 | 3.04 |
| 13.06.2031 | 14.10.2031 | 25.05.2032 | 76.9 | 108.1 | 114.2 | 15.6 | 230.7 | 3.84 | 3.07 |
| 14.06.2031 | 14.10.2031 | 25.05.2032 | 76.1 | 109.0 | 113.7 | 16.4 | 227.1 | 3.86 | 3.10 |
| 15.06.2031 | 14.10.2031 | 25.05.2032 | 75.4 | 109.7 | 113.2 | 17.2 | 224.2 | 3.88 | 3.13 |
| 16.06.2031 | 13.10.2031 | 24.05.2032 | 74.7 | 110.4 | 112.6 | 17.8 | 221.8 | 3.90 | 3.16 |
| 17.06.2031 | 13.10.2031 | 24.05.2032 | 74.1 | 111.1 | 112.0 | 18.5 | 219.8 | 3.92 | 3.19 |
| 18.06.2031 | 13.10.2031 | 24.05.2032 | 73.5 | 111.8 | 111.5 | 19.1 | 218.3 | 3.95 | 3.22 |
| 19.06.2031 | 13.10.2031 | 24.05.2032 | 72.9 | 112.4 | 110.9 | 19.7 | 217.1 | 3.97 | 3.26 |
| 20.06.2031 | 13.10.2031 | 24.05.2032 | 72.4 | 113.0 | 110.2 | 20.3 | 216.2 | 4.01 | 3.29 |